
\overfullrule=0pt
\input amssym
\input harvmac
\input epsf


\def\a{{\alpha}}

\def\b{{\beta}}
\def\g{{\gamma}}
\def\d{{\delta}}
\def\e{{\epsilon}}
\def\s{{\sigma}}

\def\half{{1\over 2}}
\def\p{{\partial}}

\def\t{{\theta}}
\def\hat{\widehat}
\def\bar{\overline}

\def\th{{\widehat\theta}}

\def\CH{{\cal H}}
\def\CO{{\cal O}}

\def\CK{{\cal K}}
\def\CV{{\cal V}}
\def\CU{{\cal U}}
\def\ll{{\langle}}
\def\rr{{\rangle}}

\def\half{{1 \over 2}}
\def\dzm{{\partial}}

\def\p{{\partial}}
\def\s{{\sigma}}

\def\a {{\alpha}}
\def\b {{\beta}}

\def\g {{\gamma}}
\def\d {{\delta}}
\def\e {{\epsilon}}

\def\ad {{\dot\alpha}}
\def\bd {{\dot\beta}}

\def\ta {{\theta^\alpha}}

\def\ad {{\dot \a}}
\def\bd {{\dot \b}}
\def\tb {{\overline\theta}}

\def\CT{{\cal T}}
\def\CG{{\cal G}}
\def\CJ{{\cal J}}
\def\CD{{\cal D}}
\def\ll{{\langle}}
\def\rr{{\rangle}}
\def\db{{\bar d}}

\def\a{\alpha}
\def\b{\beta}
\def\d{\delta}

\def\th{\theta}
\def\g{\gamma}

\def\e{\epsilon}

\def\s{\sigma}


\def\sqr#1#2{{\vbox{\hrule height.#2pt\hbox{\vrule width
.#2pt height#1pt \kern#1pt\vrule width.#2pt}\hrule height.#2pt}}}
\def\Box{\mathchoice\sqr64\sqr64\sqr{4.2}3\sqr33}


\lref\berk{N. Berkovits, {\it Super-Poincar\'e
Covariant Quantization of the Superstring,} JHEP 04 (2000) 018,
[arXiv:hep-th/0001035].}

\lref\adswitten{N. Berkovits, C. Vafa and E. Witten,
{\it Conformal Field Theory
of $AdS$ Background with Ramond-Ramond Flux,} JHEP 9903 (1999)
018, [arXiv:hep-th/9902098].}

\lref\bz{N. Berkovits, M. Bershadsky, T. Hauer, S. Zhukov and B. Zwiebach,
{\it Superstring Theory on $AdS_2\times S^2$ as a Coset Supermanifold,}
Nucl. Phys. B567 (2000) 61, [arXiv:hep-th/9907200].}

\lref\six{N. Berkovits, {\it Quantization of the
Type II Superstring in a Curved Six-Dimensional Background,}
Nucl. Phys. B565 (2000) 333, [arXiv:hep-th/9908041].}

\lref\fields{W. Siegel, {\it Fields,} hep-th/9912205.}

\lref\twohyb{N. Berkovits, S. Gukov and B. C. Vallilo, {\it 
Superstrings in 2D backgrounds with R-R flux and new extremal blackholes,}
  Nucl.\ Phys.\ B {\bf 614}, 195 (2001), [arXiv:hep-th/0107140].}

\lref\Nat{N. Berkovits, {\it Covariant Quantization of the
Green-Schwarz Superstring in a Calabi-Yau Background}, Nucl. Phys. B431,
258 (1994), [arXiv:hep-th/9404162].}

\lref\NV{N. Berkovits and C. Vafa, {\it $N=4$ Topological Strings}
, Nucl. Phys. B433 (1995) 123, [arXiv:hep-th/9407190].}

\lref\Eff{N. Berkovits and W. Siegel,
{\it Superspace Effective Actions for 4D Compactifications of Heterotic
and Type II Superstrings,} Nucl. Phys. B462 (1996) 213, 
[arXiv:hep-th/9510106].}

\lref\CVsigma{S.~Cecotti and C.~Vafa, ``Exact Results for Supersymmetric
Sigma Models", Phys.Rev.Lett. {\bf 68} (1992) 903.}

\lref\NVW{N. Berkovits, C. Vafa and E. Witten, ``Conformal Field Theory of
AdS Background with Ramond-Ramond Flux", JHEP {\bf 9903} (1999) 018,
hep-th/9902098.}

\lref\OOY{H.~Ooguri, Y.~Oz and Z.~Yin, ``D-Branes on Calabi-Yau
Spaces and Their Mirrors", Nucl.Phys. {\bf B477} (1996) 407.}

\lref\LVW{W.~Lerche, C.~Vafa and N.P.~Warner, ``Chiral Rings in
N=2 Superconformal Theories", Nucl.Phys. {\bf B324} (1989) 427.}

\lref\BCOV{M.~ Bershadsky, S.~ Cecotti, H.~ Ooguri and C.~ Vafa,
``Kodaira-Spencer Theory of Gravity and Exact Results for
Quantum String Amplitudes," Commun.Math.Phys. {\bf 165} (1994) 311.}

\lref\pureconf{N. Berkovits,{\it Pure spinor formalism as an N = 2 
topological string,} JHEP {\bf 0510}, 089 (2005), [arXiv:hep-th/0509120].}

\lref\shava{
  S.~L.~Shatashvili and C.~Vafa,
  {\it Superstrings and manifold of exceptional holonomy,}
  Selecta Math.\  {\bf 1}, 347 (1995)
  [arXiv:hep-th/9407025].
}

\lref\BuchbinderQV{
  I.~L.~Buchbinder and S.~M.~Kuzenko,
  {\it Ideas and methods of supersymmetry and supergravity: Or a walk through
  superspace.}}

\lref\WessCP{J.~Wess and J.~Bagger,{\it Supersymmetry and supergravity.}}

\lref\BandW{Wm D Linch {{III}} and Brenno Carlini Vallilo, 
Unpublished private secret communication.}

\lref\gpone{K.~Lee and W.~Siegel,
  {\it Conquest of the ghost pyramid of the superstring,}
  JHEP {\bf 0508}, 102 (2005)
  [arXiv:hep-th/0506198].
}
\lref\gptwo{K.~Lee and W.~Siegel,
  {\it Simpler superstring scattering,}
  arXiv:hep-th/0603218.
}

\lref\grana{
  M.~Grana, {\it Flux compactifications in string theory: 
A comprehensive review,}
  Phys.\ Rept.\  {\bf 423}, 91 (2006)
  [arXiv:hep-th/0509003].
}

\lref\BerkovitsNR{
  N.~Berkovits,
  {\it Review of open superstring field theory,}
  arXiv:hep-th/0105230.
}

\lref\BerkovitsIM{
  N.~Berkovits,
  {\it The Ramond sector of open superstring field theory,}
  JHEP {\bf 0111}, 047 (2001)
  [arXiv:hep-th/0109100].
}

\lref\GreeneCY{
  B.~R.~Greene,
  {\it String theory on Calabi-Yau manifolds,}
  arXiv:hep-th/9702155.
}

\lref\DixonBG{For a review, see 
  L.~J.~Dixon, {\it Some World Sheet Properties Of Superstring 
Compactifications, On Orbifolds And Otherwise,}
PUPT-1074, Lectures given at the 1987 ICTP Summer Workshop in High Energy 
Phsyics and Cosmology, 
Trieste, Italy, Jun 29 - Aug 7, 1987; J.~Distler, {\it Notes on N=2 sigma 
models,} arXiv:hep-th/9212062.
}

\lref\SeibergYZ{
  N.~Seiberg,
  {\it Noncommutative superspace, N = 1/2 supersymmetry, field theory 
and  string theory,}
  JHEP {\bf 0306}, 010 (2003)
  [arXiv:hep-th/0305248].
}

\lref\BerkovitsPQ{
  N.~Berkovits, H.~Ooguri and C.~Vafa,
  {\it On the worldsheet derivation of large N dualities for the superstring,}
  Commun.\ Math.\ Phys.\  {\bf 252}, 259 (2004)
  [arXiv:hep-th/0310118].
}

\lref\OoguriQP{
  H.~Ooguri and C.~Vafa,
  {\it The C-deformation of gluino and non-planar diagrams,}
  Adv.\ Theor.\ Math.\ Phys.\  {\bf 7}, 53 (2003)
  [arXiv:hep-th/0302109].
}

\lref\OoguriGX{
  H.~Ooguri and C.~Vafa,
  {\it Worldsheet derivation of a large N duality,}
  Nucl.\ Phys.\ B {\bf 641}, 3 (2002)
  [arXiv:hep-th/0205297].
}

\lref\BerkovitsZQ{
  N.~Berkovits, M.~Bershadsky, T.~Hauer, S.~Zhukov and B.~Zwiebach,
  {\it Superstring theory on AdS(2) x S(2) as a coset supermanifold,}
  Nucl.\ Phys.\ B {\bf 567}, 61 (2000)
  [arXiv:hep-th/9907200].
}

\lref\GiddingsYU{
  S.~B.~Giddings, S.~Kachru and J.~Polchinski,
  {\it Hierarchies from fluxes in string compactifications,}
  Phys.\ Rev.\ D {\bf 66}, 106006 (2002)
  [arXiv:hep-th/0105097].
}

\lref\LawrenceZK{
  A.~Lawrence and J.~McGreevy,
  {\it Local string models of soft supersymmetry breaking,}
  JHEP {\bf 0406}, 007 (2004)
  [arXiv:hep-th/0401034].
}

\lref\deBoerPT{
  J.~de Boer, A.~Naqvi and A.~Shomer,
  {\it The topological G(2) string,}
  arXiv:hep-th/0506211.
}

\lref\AnguelovaCV{
  L.~Anguelova, P.~de Medeiros and A.~Sinkovics,
  {\it Topological membrane theory from Mathai-Quillen formalism,}
  arXiv:hep-th/0507089.
}

\lref\AcharyaGB{
  B.~S.~Acharya,
  {\it On realising N = 1 super Yang-Mills in M theory,}
  arXiv:hep-th/0011089.
}

\lref\AcharyaQE{
  B.~S.~Acharya and S.~Gukov,
  {\it M theory and Singularities of Exceptional Holonomy Manifolds,}
  Phys.\ Rept.\  {\bf 392}, 121 (2004)
  [arXiv:hep-th/0409191].
}

\lref\TaylorII{
  T.~R.~Taylor and C.~Vafa,
  {\it RR flux on Calabi-Yau and partial supersymmetry breaking,}
  Phys.\ Lett.\ B {\bf 474}, 130 (2000)
  [arXiv:hep-th/9912152].
}

\lref\RocekIJ{
  M.~Rocek, C.~Vafa and S.~Vandoren,
  {\it Hypermultiplets and topological strings,}
  JHEP {\bf 0602}, 062 (2006)
  [arXiv:hep-th/0512206].
}

\lref\VafaWI{
  C.~Vafa,
  {\it Superstrings and topological strings at large N,}
  J.\ Math.\ Phys.\  {\bf 42}, 2798 (2001)
  [arXiv:hep-th/0008142].
}

\lref\GukovGR{
  S.~Gukov,
  {\it Solitons, superpotentials and calibrations,}
  Nucl.\ Phys.\ B {\bf 574}, 169 (2000)
  [arXiv:hep-th/9911011].
}

\lref\GukovYA{
  S.~Gukov, C.~Vafa and E.~Witten,
  {\it CFTs from Calabi-Yau four-folds,}
  Nucl.\ Phys.\ B {\bf 584}, 69 (2000)
  [Erratum-ibid.\ B {\bf 608}, 477 (2001)]
  [arXiv:hep-th/9906070].
}

\lref\KachruJE{
  S.~Kachru and J.~McGreevy,
  {\it M-theory on manifolds of G(2) holonomy and type IIA orientifolds,}
  JHEP {\bf 0106}, 027 (2001)
  [arXiv:hep-th/0103223].
}

\lref\joyce{ D. Joyce, {\it Bla....}}

\lref\BlumenhagenJB{
  R.~Blumenhagen and V.~Braun,
  {\it Superconformal field theories for compact G(2) manifolds,}
  JHEP {\bf 0112}, 006 (2001)
  [arXiv:hep-th/0110232].
}

\lref\DijkgraafTE{
  R.~Dijkgraaf, S.~Gukov, A.~Neitzke and C.~Vafa,
  {\it Topological M-theory as unification of form theories of gravity,}
  arXiv:hep-th/0411073.
}

\lref\GukovYA{
  S.~Gukov, C.~Vafa and E.~Witten,
  {\it CFTs from Calabi-Yau four-folds,}
  Nucl.\ Phys.\ B {\bf 584}, 69 (2000)
  [Erratum-ibid.\ B {\bf 608}, 477 (2001)]
  [arXiv:hep-th/9906070].
}

\lref\ArkaniHamedTB{
  N.~Arkani-Hamed, T.~Gregoire and J.~Wacker,
  {\it Higher dimensional supersymmetry in 4D superspace,}
  JHEP {\bf 0203}, 055 (2002)
  [arXiv:hep-th/0101233].
}
\lref\MSS{
  N.~Marcus, A.~Sagnotti and W.~Siegel,
{\it Ten-Dimensional Supersymmetric Yang-Mills Theory In Terms Of
Four-Dimensional Superfields,}
  Nucl.\ Phys.\ B {\bf 224}, 159 (1983).}

\lref\koba{
  N.~Berkovits,
  {\it Super-Poincare Invariant Koba-Nielsen Formulas for the Superstring,}
  Phys.\ Lett.\ B {\bf 385}, 109 (1996)
  [arXiv:hep-th/9604120].
}

\lref\onehyb{
  N.~Berkovits and B.~C.~Vallilo,
 {\it One loop N-point superstring amplitudes with manifest d = 4
  supersymmetry,}
  Nucl.\ Phys.\ B {\bf 624}, 45 (2002)
  [arXiv:hep-th/0110168].
}

\lref\PapadopoulosDA{
  G.~Papadopoulos and P.~K.~Townsend,
  {\it Compactification of D = 11 supergravity on spaces of exceptional
  holonomy,}
  Phys.\ Lett.\ B {\bf 357}, 300 (1995)
  [arXiv:hep-th/9506150].
}
\lref\DijkgraafTE{
  R.~Dijkgraaf, S.~Gukov, A.~Neitzke and C.~Vafa,
   {\it Topological M-theory as unification of form theories of gravity,}
  arXiv:hep-th/0411073.
}

\lref\GualtieriDX{ M.~Gualtieri, {\it Generalized complex geometry,} 
arXiv:math.dg/0401221.}

\lref\SiegelPX{W.~Siegel, {\it Curved Extended Superspace from Yang-Mills 
Theory a la Strings,} Phys.\ Rev.\ D {\bf 53}, 3324 (1996) 
[arXiv:hep-th/9510150].}

\lref\roiban{
  R.~Roiban and J.~Walcher,
  ``Rational Conformal Field Theories With G(2) Holonomy,''
  JHEP {\bf 0112}, 008 (2001)
  [arXiv:hep-th/0110302].
}


\Title{ \vbox{\baselineskip12pt
\hbox{YITP-SB-06-30}}}
{{\vbox{\centerline{Hybrid Formalism, Supersymmetry Reduction,}
\vskip .1in
\centerline{and Ramond-Ramond Fluxes}
}}}
\centerline{William D. Linch {III}${}^{{\spadesuit},{\clubsuit},}$\foot{E-Mail: 
wdlinch3 (at) math.sunysb.edu} and 
Brenno Carlini Vallilo${}^{\spadesuit,}$\foot{E-Mail: vallilo (at) 
insti.physics.sunysb.edu}}
\bigskip
\centerline{${}^\spadesuit${\it C.N. Yang Institute for Theoretical 
Physics} and ${}^\clubsuit${\it Department of Mathematics}}
\centerline{{\it SUNY, Stony Brook, NY 11794-3840, USA}}

\vskip .3in

The supersymmetric hybrid formalism for Type II strings is used 
to study partial supersymmetry breaking in four and three dimensions. 
We use worldsheet techniques to derive effects of internal Ramond-Ramond 
fluxes such as torsions, superpotentials and warping. 

\Date{July 2006}

\listtoc
\writetoc

\newsec{Introduction}

String theory in background fields, especially Ramond-Ramond backgrounds, 
is one of the main themes in the field. It has become clear that in order 
to gain a better understanding of many phenomena of recent interest 
we cannot avoid studying RR flux in the stringy regime.

For more than ten years,
Berkovits and collaborators have been developing a 
series of super-Poincar\'e covariant formalisms for the superstring. 
In two \twohyb, four \Nat, and six \refs{\NV,\six}\ dimensions the hybrid 
formalism is 
obtained from a field redefinition of the RNS superstring and has 
an enhanced symmetry algebra on the worldsheet, namely $N=4$, $\hat c=2$ 
superconformal symmetry. Although they share a common structure,
these algebras 
have different expressions in terms of the fundamental fields in each 
dimension and all of them have chiral bosons as fundamental worldsheet 
fields. It is natural to ask whether there is a general principle behind these 
formulations other than the underlying relation with the RNS 
superstring. 

The minimal\foot{Here we are using the nomenclature introduced in \pureconf.} 
pure spinor formalism \berk\ holds a special place among all these 
formalisms;  
it has no superconformal symmetry and no scalar ghosts. Instead, it has a set 
bosonic ghosts in the spinor representation of the Lorenz group satisfying 
the pure spinor constraint. Similarly, in the covariant approach \six\ for the 
six-dimensional hybrid formalism one is forced\foot{This procedure was 
described in footnotes in \berk.} to introduce a set of unconstrained spinor 
ghosts and a BRST charge $Q=\oint u^\alpha \CD_\alpha$ where 
$\CD_\alpha$ is the projective superspace derivative. These new ghosts 
and BRST charge have nothing to do with the underlying RNS formalism and 
are required to have manifest supersymmetry in $d=6$. In this case the 
bosonic spinor ghosts do not have to satisfy any constraint since the 
projective superspace derivative commutes with itself. (If one tries 
to increase the number of supersymmetries again one finds that it is not 
possible to construct a set of commuting supersymmetric derivatives, 
and thus the 
pure spinor constraint is necessary.) Nevertheless these 
new ghosts interact with the original variables and must appear in the 
vertex operators. We note that this case straddles the usual hybrid 
formalism and the pure spinor superstring. It would be very interesting to 
find a deeper relation between them, possibly through the superconformal 
extension of the minimal pure spinor \pureconf. 

Although supersymmetry is one of the main ingredients in string theory, 
superspace techniques have historically always played a peripheral role. 
The 
first reason is that until recently there was no quantizable 
formalism for the superstring with all supersymmetries manifest.\foot{In two 
interesting papers Lee and Siegel \refs{\gpone,\gptwo}\ introduced a new 
formalism for the superstring based on the usual GS formalism but with a 
consistent BRST charge 
built on an 
infinite pyramid of ghosts in addition to the usual $(b,c)$ system of 
the bosonic string. Scattering 
amplitudes are computed using very simple rules and its application to 
other problems seems promising.}
The second reason is that many interesting phenomena in string theory 
(especially the construction of models which resemble the observed 
particle physics) appear after breaking some supersymmetry. For example, 
ten-dimensional 
Type I superspace has encoded within itself the $N=1$ four-dimensional 
superspace but from the ten-dimensional point of view it is difficult to 
see how holomorphicity and non-renormalization theorems appear.  This has 
become 
clear in superstring field theory \refs{\BerkovitsNR,\BerkovitsIM}\ where 
chiral and 
anti-chiral $F$-terms appear, but the procedure is only possible after breaking 
some manifest Lorentz
symmetry. Finally, one can 
argue that the usefulness of superspace in higher dimensions is restricted
by our lack of understanding of it's off-shell structure. 

Breaking supersymmetry in higher-dimensional theories is usually done in 
components and the superspace, if it is introduced at all, only appears 
in the very beginning and end of the analysis. Furthermore, the Grassmann 
coordinates related 
to higher supersymmetries are simply set to zero by hand. In a supersymmetric 
covariant 
formalism it is not consistent to do this since the superspace coordinates 
are part of the conformal field theory describing the superstring; they
have to satisfy consistency conditions like vanishing of central charge 
and are essential for the worldsheet symmetries. In order to study 
compactifications of covariant 
formalisms we need a method to dimensionally reduce and break supersymmetry 
keeping all of the original superspace coordinates.\foot{The 
inverse problem, i.e. the use of standard four-dimensional $N=1$ 
superspace to describe higher-dimensional theories and those with more 
supersymmetry, has been 
widely applied in the literature. For a relatively recent discussion with 
applications to phenomenology see \ArkaniHamedTB. } In this paper we take a 
small step in this direction by studying the standard four-dimensional 
hybrid superstring in backgrounds with reduced symmetry such as
Type IIA on a $G_2$ holonomy manifold and with Ramond-Ramond flux in the 
internal space. 

The hybrid formalism has been used previously to study strings in RR 
backgrounds, for example, 
in the case of $AdS_3\times S^3$ \adswitten, $AdS_2\times S^2$ \BerkovitsZQ, 
the $C$-deformation \OoguriQP\ and 
noncommutative superspace \SeibergYZ. In all of these examples the RR flux 
considered was in the uncompactified sector of space-time, with no 
contributions coming from the internal manifold. The relation of internal 
RR fluxes with auxiliary fields in the four-dimensional supersymmetric 
multiplets first appeared in \VafaWI\ and was discussed further in 
\BerkovitsPQ\ and \LawrenceZK. Compactifications including fluxes 
have attracted a lot of attention in recent years due to 
their applications to the problem of moduli stabilization 
(see \grana\ and references therein). 

We hope that the superspace reduction introduced in this paper can be 
extended to the higher-dimensional versions of the hybrid formalism and 
possibly to the (minimal) pure spinor formalism. This type of superspace 
reduction 
could also be useful for finding the relation with the pure spinor superstring;
there is the possibility that the pure spinor formalism is the generating 
formalism for all covariant formalisms in lower dimensions. This, in turn, 
might help 
to better understand the new superconformal description of \pureconf.

Another application of the present work is to to study 
$G_2$ holonomy compactifications, at least in the case where the 
$G_2$ is of the form $({\rm CY}_3\times S^1)/\Bbb Z_2$. If we start with 
M-theory we have a four-dimensional 
effective 
field theory with $N=1$ supersymmetry and with the appropriate manifold 
we can obtain $N=1$ super YM \AcharyaGB . Since we do not have a (covariant) 
microscopic description of M-theory, we can further compactify the theory 
on a circle and use the duality with Type IIA to address stringy questions. 
Yet another application is the conjectured relation between Hitchin 
functionals in seven dimensions and topological strings on Calabi-Yau 
manifolds \DijkgraafTE . Since, in the hybrid formalism, there is a 
well-defined way to 
compute 
general supersymmetric amplitudes, it is possible to use the covariant 
description of the present paper to calculate amplitudes in backgrounds of 
the form $({\rm CY}_3\times {S^1})/{\Bbb Z}_2$ and see what terms 
topological amplitudes are computing in the three-dimensional 
effective action.

This paper is organized as follows. In the next section we will review the 
four-dimensional hybrid formalism, stressing features which are going to 
be useful in the subsequent sections and have not heretofore appeared in the 
literature, such as alternative descriptions of vertex operators and 
amplitudes in the compactification-dependent sector. 
The three-dimensional $N=4$, 2 and 1 
superspace  will be described in section 3. A convenient way to reduce 
the number of supersymmetries without leaving the original 
superspace will be 
introduced. Multiplets with various amounts of supersymmetry will be described.
We then show how the hybrid superstring in four dimensions 
can be used to describe string theories in lower dimensions without changing 
the number of fundamental fields in the formalism. We then apply 
these methods to give a supersymmetric description of Type IIA 
on $({\rm CY}_3\times {S^1})/{\Bbb Z}_2$. In section 4, we
discuss effects of Ramond-Ramond fields in four dimensions 
(which is easily adapted to the case of three 
dimensions using the results of section 3). In the 
concluding section we summarize the work and comment on future applications to 
problems of current interest. Finally, we include for completeness an 
appendix with the hybrid formalism reduced on $S^1$ in the conventions of section 3.

\newsec{Hybrid Formalism in $d=4$}

In this section we review basic aspects of the hybrid formalism in 
$d=4$ \Nat. Besides setting up definitions and notations, we comment on 
aspects of the formalism which have not appeared previously in the literature
such as supersymmetric amplitudes in the compactification sector.

\subsec{Action and Symmetries}
The original formulation of the hybrid superstring is as a field redefinition
of the RNS variables compactified on a Calabi-Yau background \Nat. In its 
final form there is a complete decoupling between the four-dimensional flat 
space
and the Calabi-Yau background.  
The fundamental variables of the $d=4$ (closed string) hybrid formalism 
are the $N=2$
superspace coordinates $(x, \th_L, \bar\th_L,\th_R,\bar\th_R)$, the 
conjugate momenta for the fermionic coordinates 
$(p_L,\bar p_L,p_R,\bar p_R)$ and two chiral bosons $(\rho_L,\rho_R)$.

The action is 
\eqn\hybact{ S_{\rm hybrid}=\int d^2z [ \p_L x^m\p_R x_m + p_{L\a} \p_R 
\th^\a_L + 
\bar p_{L\ad} \p_R \bar\th^\ad_L +p_{R\a} \p_L \th^\a_R + 
\bar p_{R\ad} \p_L \bar\th^\ad_R] + S_{\rm chiral},}
where $S_{\rm chiral}$ is the action for the chiral bosons. The fundamental 
OPE's are:
\eqn\OPES{x^m(z)x^n(w) \to \eta^{mn}ln|z-w|^2, }
$$p_{L\a}(y)\theta_L^\b (z)\to {\delta_\a^\b\over{y -z}},\quad
\bar p_{L\ad}(y)\bar\theta_L^\bd (z)\to {\delta_\ad^\bd\over{y -z}},$$
$$p_{R\a}(y)\theta_R^\b (z)\to {\delta_\a^\b\over{\bar y -\bar z}},\quad
\bar p_{R\ad}(y)\bar\theta_R^\bd (z)\to 
{\delta_\ad^\bd\over{\bar y -\bar z}},$$
$$ \rho_L(z)\rho_L(w)\to -ln(z-w),\quad 
\rho_R(z)\rho_R(w)\to -ln(\bar z-\bar w).$$
The last line shows that the chiral bosons are time-like and hence 
cannot be fermionized. Furthermore, they are space-time scalars. 

The action \hybact\ is supersymmetric, and the corresponding supercharges are 
\eqn\SUSY{q_{L\a}=\oint dz [p_{_L\a} -{i\over 2}
\tb_L^\ad\dzm_L x_{\a\ad}-{1\over {8}}(\tb_L)^2\dzm_L\th_{L\a}],}
$$ \bar q_{L\ad}=\oint dz [\bar p_{L\ad}
-{i\over 2}
\ta_L\dzm_L x_{\a\ad}-{1\over {8}}(\th_L)^2\dzm_L\tb_{_L\ad}],$$
$$q_{R\a}=\oint dz [p_{R\a} -{i\over 2}
\tb_R^\ad\dzm_R x_{\a\ad}-{1\over {8}}(\tb_R)^2\dzm_R\th_{R\a}],$$
$$\bar q_{R\ad}=\oint dz [\bar p_{R\ad}
-{i\over 2}
\ta_R\dzm_R x_{\a\ad}-{1\over {8}}(\th_R)^2\dzm_R\tb_{_R\ad}].$$

There is a set of operators which commutes with the charges \SUSY\ (and 
with their right-moving counterparts):
\eqn\SOPRT{d_{L\a}=p_{L\a}+{i\over 2}\tb_L^\ad\dzm x_{\a\ad}-
{1\over 4}(\tb_L)^2\dzm\th_{L\a}
+{1\over {8}}\th_{L\a} \dzm (\tb_L)^2 ,}
$$\bar d_{L\ad}=\bar p_{L\ad}
+{i\over 2}\th_L^\a\p x_{\a\ad}-{1\over 4}(\th_L)^2\p\tb_{L\ad}
+{1\over {8}}\tb_{L\ad} \dzm (\th_L)^2 ,$$
$$\Pi^m_L =\dzm_L x^m -{i\over 2}\s^m_{\a\ad}(\th_L^\a \dzm_L\tb_L^\ad+
\tb_L^\ad\dzm_L\th_L^\a) , $$
and similarly for the right moving sector. Here $x_{\a\ad}=x_m 
\sigma_{\a\ad}^m$.
These operators realize the following algebra
\eqn\dope{d_{L\a} (y) \bar d_{L\ad}(z) \to i{{\Pi_{L\a\ad}}\over{y -z}}, \quad
d_{L\a}(y) d_{L\b}(z) \to {\rm regular},\quad 
\bar d_{L\ad}(y) \bar d_{L\bd}(z) \to {\rm regular},}
$$ d_{L\a}(y) \p_L\th_L^\b(z) \to {{\d_\a^\b}\over{(y-z)^2}},
\quad
\bar d_{L\ad}(y) \p_L\tb_L^\bd(z) \to {{\d_\ad^\bd}\over{(y-z)^2}},
$$
$$
d_{L\a}(y) \Pi_L^m(z) \to -i{{\sigma^m_{\a\ad} \dzm_L\tb_L^\ad}\over{y -z}},
\quad
\bar d_{L\ad}(y) 
\Pi_L^m(z) \to -i{{\sigma^m_{\a\ad} \dzm_L\th_L^\a}\over{y -z}},$$
$$\Pi_L^m(z)\Pi_L^n(z)\to -{{\eta^{mn}}\over{(y-z)^2}}.$$

Although it is not manifest, the action \hybact\ in invariant under a 
non-linear $N=(2,2)$ 
superconformal transformation which is generated by
\eqn\SUAL{T_L=-\half\p_L x^m \p_L x_m -
p_{L\a}\p_L \th_L^\a - \bar p_{L\ad }\p_L\tb_L^\ad -\half\p_L\rho_L\p_L\rho_L,}
$$G_L^+=e^{\rho_L} (d_L)^2 , \quad
G_L^-=e^{-\rho_L} (\bar d_L)^2, \quad
J_L=-\dzm_L\rho_L,
$$
again, together with the right-moving counterpart.

\subsec{Coupling to $c=9$, $N=2$ CFTs}

We can couple this $c=-3$, $N=2$ CFT to 
any $N=2$ CFT. 
Consistency of an $N=2$ superstring theory requires that the total 
central charge be $c=6$. This 
is the familiar
condition for 
the standard critical $N=2$ string, after the introduction of the ghost 
sector. In the hybrid formalism this condition is 
better seen as a requirement to admit an $N=4$ topological description \NV\ 
in which no additional superconformal ghosts are needed. 
The $N=4$ formalism is suitable for defining 
scattering amplitudes and a string field theory action in the case of open 
strings. We will introduce some of its properties when needed.

There is the possibility of coupling the hybrid variables to a more 
supersymmetric CFT. This means that there are some space-time 
supersymmetries that are not linearly realized in the fundamental 
variables, and the hybrid description is not the most economical. The trivial 
example is the six-dimensional torus. Nevertheless, this example is very 
useful to compute exact answers in the CFT.

Given a $c=9$, $N=2$ CFT with left-moving generators 
$(\CT_L,\CG_L^+,\CG_L^-,\CJ_L)$ a consistent 
string theory has action
$$S = S_{hybrid}+S_{c=9},$$
and $c=6$, $N=2$ generators
\eqn\fullscft{T_L= -\half\p_L x^m \p_L x_m -
p_{L\a}\p_L \th_L^\a - \bar p_{L\ad }\p_L\tb_L^\ad -\half\p_L\rho_L\p_L\rho_L
  + \CT_L,}
$$G^+_L= e^{\rho_L} (d_L)^2 + \CG^+_L,\quad G^-_L=e^{-\rho_L} (\bar d_L)^2  
+\CG^-_L,$$
$$J_L=  -\dzm_L\rho_L + \CJ_L.$$
The right-moving sector of the algebra is determined by the 
choice of Type IIA or Type IIB superstring. 
A consistent convention
for the present work is the following\foot{ This differs from the conventions 
in \Eff.}
\eqn\iiachoice{{\rm Type\, IIA:}\;\; G^+_R= e^{\rho_R}(\bar d_R)^2+\CG^-_R,
\quad 
G^-_R=e^{-\rho_R}(d_R)^2+\CG^+_R,\quad J_R=-\p_R\rho_R -\CJ_R}
\eqn\iibchoice{{\rm Type\, IIB:}\;\; G^+_R=e^{\rho_R}(d_R)^2+\CG^-_R,\quad 
G^-_R=e^{-\rho_R}(\bar d_R)^2+\CG^+_R,\quad J_R=-\p_R\rho_R -\CJ_R }
The change in the energy-momentum tensor is given simply by switching 
$L\to R$ in derivatives and fields.

Since now the central charge is $c=6$, this system defines a critical 
$N=2$ superstring. One could now add superconformal ghosts and perform 
standard BRST quantization to define physical states and amplitudes 
but we will show that this is not necessary.

In the superconformal field theory of a Calabi-Yau 3-fold background 
we have, in addition to 
the usual superconformal algebra, a {\it second} superconformal 
algebra that does {\it not} commute with the first. We will call 
these generators 
\eqn\secondal{\tilde t_L={1\over 6}\CJ_L^2,\quad\tilde g^+_L= 
{1\over \sqrt{3}} \Omega^+_L=e^{\CH_L},\quad \tilde g^-_L=
{1\over \sqrt{3}}\Omega^-_L={1\over \sqrt{3}}e^{-\CH_L},\quad 
\tilde j={1\over 3}\CJ_L,} 
where $\p_L\CH_L=\CJ_L$ and $\Omega^+_L$ and $\Omega^-_L$ are holomorphic 
chiral and anti-chiral fields with charge $3$ and $-3$ of the original 
superconformal algebra. They can be 
written in terms of the holomorphic $\Omega_{IJK}$ and anti-holomorphic 
$\bar\Omega_{\bar I\bar J\bar K}$ 3-form on the 
Calabi-Yau respectively. The same applies to the right-moving sector and together these 
operators form an important part of the $N=4$ superconformal algebra. 

To construct the extended superconformal algebra \NV , we note that 
\eqn\sutwoalgebra{ J_L=-\dzm_L\rho_L + \CJ_L, \quad J^{++}_L=e^{-\rho_L}\Omega^+_L, \quad 
J^{--}_L=e^{+\rho_L}\Omega^-_L,}
form an $su(2)$ current algebra. With these operators we can generate 
two new superconformal operators
\eqn\newscft{\tilde G^+_L= J^{++}_L(G^-_L)=
\Omega^+_L e^{-2\rho_L}(\bar d_L)^2+ 
e^{-\rho_L}\Omega^+_L(\CG^-_L),}
$$\tilde G^-_L= J^{--}_L(G^+_L)= \Omega^-_L e^{2\rho_L}(d_L)^2+
e^{\rho_L}\Omega^-_L(G^+_L), $$
The action of 
$(J^{++}_L, J^{--}_L)$ on all other supercharges vanishes. A 
similar 
construction works in the right moving sector, but one should mind 
the conventions expressed in \iiachoice\ and \iibchoice. The constraints 
$\{ T_{L,R}, G^\pm_{L,R}, \tilde G^\pm_{L,R}, J_{L,R}, J_{L,R}^{\pm\pm}\}$ 
generate the required $N=4$ algebra.

\subsec{Physical States}

We define physical states as primary fields of the algebra \fullscft. Due 
to 
the large worldsheet symmetry algebra, vertex operators can be written in many 
equivalent ways. For every physical state, there is an infinite number of
vertex operators representing it \NV. This large degeneracy is reminiscent of 
picture changing in the RNS formalism.  
Depending on the application, some choices have proven to be  more useful 
than others. We are going 
to introduce some of them, explaining when each choice is suitable. 

Massless vertex operators are constructed using neutral operators of conformal 
dimension zero times 
a function of the zero-modes of $(x^m, \t_L^\a, \tb_L^\ad, \t_R^\ad, 
\tb_R^\ad)$. In the simplest case the operator of conformal dimension zero 
is the identity operator $1$, and the vertex operator is just 
$V=U(x^m, \t_L^\a, \tb_L^\ad, \t_R^\ad, \tb_R^\ad)\times 1$. $V$ is a primary
field of conformal weight zero if 
$$(T_L)_0 V=(G^+_L)_\half V=(G^-_L)_\half V=0,$$
together with the right-moving counterpart. Here $\CO_n \CA$ means the 
pole of order $h_{\CO}+n$ in the OPE of 
$\CO$ and $\CA$ where $h_\CO$ is the conformal weight of $\CO$. Using the algebra \fullscft\ we have 
\eqn\poleq{\nabla_L^2 U=\bar\nabla_L^2 U=\nabla_R^2 U=\bar\nabla_R^2 U= 
\Box_4 U=0, }
where the $\nabla$ are the superspace covariant derivatives. These equations 
imply polarization 
and mass shell conditions for the superfield $U$ and it can be shown
that $U$ is the prepotential for $N=2$ supergravity plus a tensor 
multiplet in a supersymmetric gauge. It is a general feature of the hybrid formalism that  prepotentials 
(vs. potential or strength superfields) appear in 
the unintegrated vertex operators. 

In principle one could also consider
$U$ to be a real function of the Calabi-Yau coordinates 
$(y^I,\bar y^{\bar I})$. 
We will require that this function is smooth so that it does not depend on the cohomology of the ${\rm CY}_3$. 
We will see that if there are NSNS and RR fluxes in the Calabi-Yau, $U$ will have such non-trivial $(y^I,\bar y^{\bar I})$ dependence.

The other basic primary fields for Type II strings come from chiral and 
twisted-chiral 
operators in the Calabi-Yau CFT \DixonBG. These operators are classified 
by their charges $(q_L,q_R)$ under $(\CJ_L,\CJ_R)$. Since they are 
(anti-)chiral their conformal weight 
is determined by their charge as $h={1\over 2}|q|$. Operators of charge $(-1,-1)$ 
are annihilated 
 by both $(\CG^-_L,\CG^-_R)$ and $(\Omega^-_L,\Omega^-_R)$ and the ones with 
charge $(-1,1)$ 
are annihilated by $(\CG^-_L,\CG^+_R)$ and $(\Omega^-_L,\Omega^+_R)$. Together 
with their complex 
conjugates, these operators describe the K\"ahler and complex compactification
moduli respectively. 
Let  $\Phi^a$ and $\Psi^i$ be the operators with charges  $(-1,-1)$ and 
$(-1,1)$ respectively. Then the 
vertex operators are given by
\eqn\phymoduli{ \Xi=M_a\Phi^a, \quad\quad  \Sigma=H_i\Psi^i,}
where $a$ runs from $1$ to $h^{1,1}$ (the number of K\"ahler parameters),
$i$ runs from $1$ to $h^{2,1}$ (the number of complex structure 
deformations), and $M_a$ and $H_i$ are space-time superfields. Since $\Xi$ 
and $\Sigma$ are charged, appropriate physical state conditions are 
\eqn\physxi{(G^+_L)_{-\half}\Xi=(G^-_R)_{-\half}\Xi=(G^-_L)_{\half}\Xi=
(G^+_R)_{\half}\Xi=0,}
\eqn\physsigma{(G^+_L)_{-\half}\Sigma=(G^+_R)_{-\half}\Sigma=
(G^-_L)_{\half}\Sigma=(G^-_R)_{\half}\Sigma=0.}

Using \fullscft\ and \iiachoice\ these conditions for Type IIA imply that 
$M_a$ is chiral superfield 
and it is physical when $\nabla^2_L M_a=\nabla^2_R M_a=0$. It therefore 
describes an $N=2$ vector superfield. Similarly,
$H_i$ is a twisted-chiral superfield which is physical when 
$\nabla^2_L H_i=\bar\nabla^2_R H_i=0$, describing an $N=2$ tensor multiplet. 
In the case of Type IIB we use \iibchoice, and the roles of $M_a$ and $H_i$ 
are reversed. The form \phymoduli\ is convenient for describing 
deformations of the hybrid string action. 
This is the usual description of the massless physical states of the theory. 
There are, however, alternative descriptions.
We will discuss this for the Type IIA case only as it can be easily modified 
for Type IIB. 

First we note that $\Xi$ and $\Sigma$ can also be described by uncharged 
operators. This comes from 
the fact that $e^{\rho_L}d^2_L$ and $e^{-\rho_L}\bar d^2_L$ together with 
their right moving 
(keeping in mind \iiachoice)
counterparts are {\it invertable}, that is
$$(e^{\rho_L}d^2_L)_{-\half}(e^{-\rho_L}\t^2_L)=1, \quad 
(e^{-\rho_L}\bar d^2_L)_{-\half}(e^{\rho_L}\bar \t^2_L)=1.$$
Using left- and right-moving combinations of $e^{-\rho_L}\t^2_L$ and 
$e^{\rho_L}\bar\t^2_L$ it is therefore
possible to write 
\eqn\uncharged{ \xi = (e^{\rho_L}\bar\t^2_L) (e^{-\rho_R}\bar\t^2_R) \Xi= 
\bar\t^2_L\bar\t^2_R M_a  e^{\rho_L-\rho_R}\Phi^a,} 
$$\sigma = (e^{\rho_L}\bar\t^2_L) (e^{\rho_R}\t^2_R) \Sigma= 
\bar\t^2_L\t^2_R H_i  e^{\rho_L+\rho_R}\Psi^i.$$

Because $\t$ appears explicitly in the equations above, $\xi$ and $\sigma$ do 
not look supersymmetric. Remembering, however, that $M_a$ and $H_i$ are chiral and twisted 
chiral allows us to 
write $M_a=\bar\nabla^2_L\bar\nabla^2_R(\bar\t^2_L\bar\t^2_R M_a)$ and 
$H_i=\bar\nabla^2_L\nabla^2_R(\bar\t^2_L\t^2_R H_i)$ or in a more general 
gauge, 
$M_a=\bar\nabla^2_L\bar\nabla^2_R m_a$ and $H_i=\bar\nabla^2_L\nabla^2_R h_i$, 
with complex unconstrained 
$m_a$ and $h_i$. The final result is that the uncharged operators can be 
written as 
\eqn\neutralmodu{\xi=m_a e^{\rho_L-\rho_R}\Phi^a, \quad \sigma=h_i 
e^{\rho_L+\rho_R}\Psi^i.}

From this one can see that $m_a$ and $h_i$ play the role of prepotentials 
for $M_a$ and $H_i$, in 
analogy with the vertex operator for the supergravity sector. Using 
\fullscft\ 
and \iiachoice\ one can show that these operators are uncharged 
with respect to the full 
superconformal algebra. The uncharged
operators are essential 
for computing scattering amplitudes.

The extended worldsheet superconformal symmetry also allows different 
descriptions of vertex operators.
Using the $su(2)$ current algebra one can transform, for example,  a chiral 
operator of the original 
superconformal algebra \fullscft\ to an anti-chiral field of 
\newscft. As will be shown below, it is useful to have operators with positive 
charge in the left-moving sector and negative charge in the right-moving one. 
The vertex operators in \phymoduli\ do not satisfy this requirement.
Applying $J^{++}_L$ on
$\Xi$ we have
$$(J^{++}_L)_0(\Xi)=M_a e^{-\rho_L}\Omega^+_L(\Phi^a),$$
where $\Omega^+_L(\Phi^a)$ is twisted-chiral primary field in the Calabi-Yau 
CFT with charge $(2,-1)$ under $(\CJ_L,\CJ_R)$. The same procedure 
is applied to $\Sigma$:
$$(J^{++}_L)_0(J^{++}_R)_0\Sigma=
H_i e^{-\rho_L-\rho_R}\Omega^+_L(\Omega^-_R(\Psi^{i})), $$
where $\Omega^+_L(\Omega^-_R(\Psi^{i}))$ has charge $(2,-2)$.
All operators generated from the actions of $\Omega^\pm_{L,M}$ on the 
original $(-1,-1)$, $(-1,1)$, $(1,1)$ and $(1,-1)$ rings can 
be organized in four different Hodge diamonds. The one with the convenient 
charges is the one shown below:

$$\matrix{ &   &         &  \Omega^+_L\Omega^-_R      &   &   &    \cr
   &   & 0       &         & 0 &   &    \cr
    & 0 && \Omega^+_L(\Omega^-_R(\Psi^{i})) &   & 0 &    \cr
  \Omega^+_L &&  \Omega^+_L(\Phi^a) && \Omega^-_R(\bar\Phi^{\bar a})   
&    &\Omega^-_R   \cr
    & 0 &         & \bar\Psi^{\bar i} &   & 0 &    \cr
   &   & 0       &         & 0 &   &    \cr
   &   &         &  1      &   &   &    }$$
\nobreak
\centerline{\vbox{\hsize=3.5 in \noindent Table 1. {\it The Hodge diamond of
Calabi-Yau threefold CFT operators with positive left-moving charge 
and negative right-moving charge.}}}
\vskip 0.3in

\subsec{Deformations of the Action}

Deformations of the action, i.e. integrated vertex operators, should 
have conformal weight $(1,1)$ and preserve 
the $N=2$ superconformal algebra. Integrated vertex operators are also 
used to compute scattering amplitudes with more than three external states. 
Starting with $U$ the only operator that satisfies these conditions is 

\eqn\udef{\delta S_U=\int d^2z |G^+_{-\half}G^-_{-\half}|^2 U}
where $|\cdot |^2$ means left- right-moving product. The explicit form of this 
vertex operator can be used to derive the full action in a general 
curved background \Eff. Compactification-dependent states can also 
be used to deform the action. These are important for the description
of general backgrounds with fluxes and warping. In the case of $M_a$, the 
vertex operator takes the form
$$\delta S_{M_a}=\int d^2z [(G^+_L)_{-\half}(G^-_R)_{-\half}\Xi +c.c.] = 
\int d^2z [|G^+_{-\half}G^-_{-\half}|^2\xi +c.c.]=$$
\eqn\defm{\int d^2z[M_a\CG^+_L( \CG^+_R(\Phi^a))+ 
e^{\rho_L-\rho_R}d^\alpha_L d^\beta_R 
(\nabla_{L\alpha} \nabla_{R\beta}M_a)\Phi^a +}
$$e^{\rho_L}d^\alpha_L(\nabla_{L\alpha}M_a)\CG^+_R(\Phi^a) +
e^{-\rho_R}d^\alpha_R(\nabla_{R\alpha}M_a)\CG^+_L(\Phi^a) + 
c.c.],$$
where $M_a=\bar\nabla_R^2 \bar\nabla_L^2 m_a$ is the chiral field strength. 
If $M_a$ is a constant superfield  
only the first term survives which corresponds to the usual result in the
RNS formalism. The other terms are required in the supersymmetric formalism 
to ensure full superconformal invariance. Deformations corresponding to 
$H_i$ can be computed similarly:
$$\delta S_{H_i}=\int d^2z [(G^+_L)_{-\half}(G^+_R)_{-\half}\Sigma +c.c.]  =
\int d^2z [|G^+_{-\half}G^-_{-\half}|^2\sigma +c.c.]$$
\eqn\defh{=\int d^2z[H_i\CG^+_L( \CG^-_R(\Psi^i))+ 
e^{\rho_L+\rho_R}d^\alpha_L \bar d^{\dot \beta}_R 
(\nabla_{L\alpha} \bar \nabla_{R\dot \beta}H_i)\Psi^i }
$$+e^{\rho_L}d^\alpha_L(\nabla_{L\alpha}H_i)\CG^-_R(\Psi^i) 
+e^{\rho_R}\bar d^{\dot\alpha}_R(\bar\nabla_{R\dot\alpha}H_i)\CG^+_L(\Psi^i) + 
c.c.]$$

\subsec{Supersymmetric Amplitudes}

Supersymmetric amplitudes can be computed in a straightforward way 
 using the rules
of the hybrid formalism \refs{\koba,\onehyb}\ in four dimensions. 
Amplitudes are calculated by twisting the superconformal algebra. This 
has the effect of shifting the conformal weights 
$(h_L,h_R)\to (h_L -{q_L\over 2}, h_R-{q_R\over 2})$ so that all operators 
defined in the zero-mode measure have conformal weight zero, as they should.
The twisting is also responsible for a charge anomaly of 2 in
the left- and right-moving sector, which is cancelled by the measure. Due to 
the charge anomaly in the algebra of \fullscft, we will have to use all of the 
operators defined above to obtain non-vanishing amplitudes.
The first step is to define the measure over zero-modes. 
To this end it should be observed that the integral of the product of the 
holomorphic and anti-holomorphic forms over the Calabi-Yau is proportional 
to the volume
$$i \int_{CY_3}\Omega \wedge \bar \Omega= {4\over 3}{\rm Vol}({\rm CY}_3) .$$
As was shown above, these forms are represented by $\Omega^+_L$ and 
$\Omega^-_R$. In the internal CFT we therefore define
\eqn\normone{\ll \Omega^+_L\Omega^-_R \rr_{{\rm CY}_3} =1}

In the non-compact sector momentum conservation is ensured by integrating 
over space-time. In addition, we have to remove the zero-modes of the 
fermionic coordinates $\t$; their conjugate momenta $p$ have no zero 
modes on the sphere. The final ingredient is the measure for 
the chiral bosons $(\rho_L,\rho_R)$. The final form of measure is\foot{Note 
that the measure is of D-term type. Since some superconformal generators 
 have trivial cohomology it is also possible to write chiral 
and twisted-chiral F-term measures \refs{\BerkovitsNR,\BerkovitsIM}.}
\eqn\normfinal{\ll \t^2_R\bar \t^2_L \t^2_R \bar\t^2_R 
e^{-\rho_L-\rho_L}\Omega^+_L\Omega^-_R \rr=1.}

The first non-vanishing amplitude is the three point function and due to 
${\rm SL}(2,{\Bbb R})$ invariance, the three vertex operators should be 
unintegrated. The charge anomaly in the $N=2$ twisted algebra factorizes 
between space-time and Calabi-Yau sectors; a $-1$ contribution should come 
from the chiral bosons and $+3$ from the (anti-)holomorphic forms. This 
requirement narrows down the possible choices. 

As a first example, let us compute the three-point function in the Type IIA 
string for the $\bar H_{\bar i}$ moduli. Two charged $\bar \Sigma$ and 
one uncharged vertex operator 
$\bar \sigma=\bar h_{\bar i} e^{-\rho_L-\rho_R}\bar \Psi^{\bar i}$ 
will be needed.
The chiral ring structure of operators in Table 1 is essential in 
the computation. It is not hard to see that the correlation between 
three $\bar\Psi$s is \DixonBG\ 
$$ \bar\Psi^{\bar i}\times \bar\Psi^{\bar j}\times \bar\Psi^{\bar k} = 
\Omega^+_L\Omega^-_R\; \CC^{\bar i\bar j\bar k}, $$
where $\CC^{\bar i\bar j\bar k}$ are the $h^{2,1}$ 
intersection numbers. The final 
answer is 
$$\ll \bar\Sigma_1 \bar\Sigma_2 \bar\sigma_3 \rr = \int d^4x 
d^2\t_L d^2\bar\t_L d^2\t_R d^2\bar\t_R 
\bar H_{\bar i}\bar H_{\bar j} \bar h_{\bar k} \CC^{\bar i\bar j\bar k},$$
where the integration over $\t$ comes from the zero-mode measure. Since 
$\bar H$'s are twisted-chiral and $\bar h_{\bar k}$ is unconstrained, 
we can further 
perform integration over $d^2\t_Ld^2\bar\t_R$ to get
$$\ll \bar\Sigma_1 \bar\Sigma_2 \bar\sigma_3 \rr=\int d^4x d^2\bar\t_L d^2\t_R 
\bar H_{\bar i}\bar H_{\bar j} \bar H_{\bar k} \CC^{\bar i\bar j\bar k},$$
which is the expected result. Similarly, it is possible to compute the 
amplitude for $H_i$ using two $(J^{++}_L)_0(J^{++}_R)_0\Sigma=
H_i e^{-\rho_L-\rho_R}\Omega^+_L(\Omega^-_R(\Psi^{i}))$ vertex operators
and one $\sigma$. The calculation is 
slightly more involved due to the correlation between the chiral bosons and 
${\rm CY}_3$ operators. A shorter path, which gives the same answer,  is 
to just take the complex conjugate 
of the previous amplitude
$$\ll [(J^{++}_L)_0(J^{++}_R)_0\Sigma_1] [(J^{++}_L)_0(J^{++}_R)_0\Sigma_2] 
\sigma_3\rr= \int d^4x d^2\t_L d^2\bar\t_R H_i H_j H_k \CC^{ijk}.$$

Now let us compute amplitudes involving $M_a$. We need two operators of the 
type $(J^{++}_L)_0(\Xi)=M_a e^{-\rho_L}\Omega^+_L(\Phi^a)$ and one $\xi$. 
This time we have
$$e^{-\rho_L}\Omega^+_L(\Phi^a)\times e^{-\rho_L}\Omega^+_L(\Phi^b) 
\times e^{\rho_L-\rho_R}\Phi^c= 
e^{-\rho_L-\rho_R}\Omega^+_L\Omega^-_R\;\CK^{abc},$$
where $\CK^{abc}$ are $h^{1,1}$ intersection numbers. Note that the factors
of $e^{\rho_L}$ and $e^{-\rho_L}$ are needed to remove poles and zeros 
in the correlators of operators in the Hodge diamond. The amplitude becomes
$$\ll [(J^{++}_L)_0\Xi_1][(J^{++}_L)_0\Xi_2]\xi_3 \rr =
\int d^4x d^2\t_L d^2\t_R M_a M_b M_c \CK^{abc}.$$
The analogous formula for anti-chiral fields is
$$\ll [(J^{++}_R)_0\bar\Xi_1][(J^{++}_R)_0\bar\Xi_2]\bar\xi_3 \rr =
\int d^4x d^2\bar\t_L d^2\bar\t_R \bar M_{\bar a} 
\bar M_{\bar b} \bar M_{\bar c}\bar\CK^{\bar a\bar b\bar c}.$$

All other three-point amplitudes involving only moduli states are zero. 
We now turn to amplitudes with more than three points. These amplitudes have 
the general form
$$\ll \CV_1\CV_2 v_3 \prod_{i=4}^{n} \int d^2z_i \CU_i \rr,$$
where $\CV$ are charged, unintegrated vertex operators, $v$ is uncharged and the
$\CU_i$ are integrated vertex operators like \defm.  All the operators used 
in the previous computations are the only ones with zero conformal weight 
in the twisted theory. Furthermore, they saturate the charge anomaly in the 
correlation functions. These two facts imply that if we consider higher 
point amplitudes, only terms like the first one in \defm\ are going to enter 
the computation, so in the hybrid formalism the usual non-renormalization 
theorems \DixonBG\ for tree level amplitudes apply. The charge anomaly 
will also be useful in Section 4, when we compute amplitudes involving 
Ramond-Ramond flux.
Four-point amplitudes can be calculated similarly, and give the special 
geometry equations relating the curvature tensor to the metric and Yukawa 
couplings (see e.g. the second reference in \DixonBG).
Since we are not going to use this type of amplitude, we will not discuss 
it further. The problem of computing one loop amplitudes as in reference  
\onehyb\ for compactification-dependent states is still open.

\newsec{Dimensional Reduction, Quotients and $G_2$ Holonomy}

In this section we discuss all multiplets defined above and introduce 
the superspace reduction that relates them to theories with less 
supersymmetry. 
We then turn to the description of the Type IIA string on 
a $({\rm CY}_3\times S^1)/{\Bbb Z}_2$ quotient with $G_2$ holonomy.

\subsec{The Superspace ${}^N\!d={}^24$ and Its Reductions}

The ${}^N\!d={}^24$ chiral $M$ and twisted-chiral $H$ field strengths are 
defined by the constraints 
$$
0=\bar \nabla_{L \dot \alpha} M=\bar \nabla_{R \dot \alpha} M~~~,~~~0=
\bar \nabla_{L \dot \alpha} H= \nabla_{R \alpha} H
$$
implying that they can be written in terms of unconstrained complex 
prepotentials $m$ and $h$
$$
M=\bar \nabla^2_L\bar \nabla^2_R m~~~,~~~H=\bar \nabla^2_L \nabla^2_R h
$$
In addition, the reality conditions
$$
0=\nabla_L^2 M-\bar \nabla^2_R \bar M~~~,~~~0= \nabla^2_L H-\nabla^2_R\bar H
$$
are imposed, resulting in the component expressions
$$
M=\varphi+\theta_L^\alpha \psi_\alpha+\theta_R^\alpha\lambda_\alpha
+\theta_L^2 F
+\theta^2_R\bar F
+\theta_L^\alpha \theta_R^\beta  \left( \varepsilon_{\alpha \beta} D+
f_{\alpha \beta}\right)+\dots
$$
$$
H=\ell+\theta_L^\alpha \eta_\alpha+
\bar \theta_{R}^{ \dot \alpha}\bar \xi_{\dot \alpha}
+\theta_L^2 y
+\bar \theta^2_R\bar y
+\theta_L^\alpha \bar \theta_R^{\dot \alpha} 
\left( i \partial_{\alpha\dot \alpha}  l 
+\tilde H_{\alpha \dot \alpha} \right)+\dots
$$
where the ellipses denote auxiliary terms. The reality conditions are 
necessary to ensure that 
$F_{mn}=(\gamma_{mn})^{\alpha \beta}f_{\alpha \beta}+{\rm h.c.}$ and 
$\tilde H_m=\epsilon_{mnpq} H^{npq}$ satisfy the appropriate Bianchi 
identities. They also put the theory partially on-shell, an inevitability 
of non-harmonic superspaces. Fortunately, this shortcoming will not 
hamper our analysis too much allowing us to avoid introducing harmonic 
superspaces in this work.

In what follows, we will make use of the dimensional reduction of this 
superspace from ${}^N\!d={}^24$ to ${}^N\!d={}^43$. We choose to single 
out $y=x_2$ for this purpose. We then use $i(\sigma^2)_{\alpha \dot \alpha}$ 
to convert all dotted spinor indices to undotted ones and define 
$(\gamma^m)_\alpha{}^\beta =i (\sigma^2 \sigma^m)_\alpha{}^\beta$ to be 
the real three-dimensional Dirac matrices. These matrices are symmetric 
upon lowering an index. The superspace coordinates can now be taken to 
be $(x^m, y, \theta_L^\alpha, \bar \theta_L^\alpha, \theta_R^\alpha, 
\bar \theta_R^\alpha)$. 
$$
\{ \nabla_{L \alpha}, \nabla_{L \beta}\}=0~~~,~~~\{ \nabla_{R \alpha}, 
\nabla_{R \beta}\}=0
$$
$$
\{ \bar \nabla_{L \alpha}, \bar \nabla_{L \beta}\}=0~~~,
~~~\{ \bar \nabla_{R \alpha}, \bar \nabla_{R \beta}\}=0
$$
$$
\{ \nabla_{L \alpha}, \bar \nabla_{L \beta}\}=
-2i\partial_{\alpha \beta}-2\varepsilon_{\alpha \beta} \partial_y~~~,
~~~\{ \nabla_{R \alpha}, \bar \nabla_{R  \beta}\}=
-2i\partial_{\alpha\beta}-2\varepsilon_{\alpha \beta} \partial_y
$$

A new feature of ${}^N\!d={}^43$ superspace is the involution exchanging 
$\theta_L^\alpha \leftrightarrow \bar\theta_L^\alpha$,
$\theta_R^\alpha \leftrightarrow \bar\theta_R^\alpha$, and taking 
$y\mapsto -y$.\foot{Starting with Type I superspace in ten dimensions and 
working 
down, one finds that
it is not possible to define an involution of this type preserving more than an
${\rm SO}(1,5)$ Lorentz symmetry. This corresponds to compactifification on 
a $K3$ surface and is a simple way to see why the first special holonomy 
manifold occurs in four dimensions.} More useful for our purposes 
is the combination of this 
involution with the usual hermitian conjugation (denoted ${}^{\bar{~}}$) 
which we will denote by $\circ$. 
Note that although $\circ$ involves the whole superspace, effectively it 
only acts on components of (twisted-)chiral superfields since the two 
involutions of which it is composed together fix $\theta_{L,R}$ and 
$\bar \theta_{L,R}$. 

Having identified a new involution it is natural to consider its 
eigenspaces. We therefore introduce new projection operators 
${\rm Re}_\circ={1\over2} (1+\circ)$ and ${\rm Im}_\circ=
{1\over 2i}(1-\circ)$ acting on superfields. Similarly to the ordinary 
real and imaginary subspaces of complexified superspace,
the $\circ$-real and $\circ$-imaginary superspaces are half-supersymmetric. 
In this way, the ${}^N\!d={}^43$ representations are reduced 
to ${}^N\!d={}^23$.

Let let $X={\rm Re}_\circ N$ and $Y={\rm Im}_\circ N$ denote 
the $\circ$-real and $\circ$-imaginary parts of a general ${}^N\!d={}^24$ 
superfield $N$. Then $N=X+i Y$. Note that $\bar X\neq X$ and similarly 
for $Y$ so these fields are not real with respect to the original 
${}^{\bar{~}}$-conjugation. However, it is easy to see that under 
left-moving supersymmetry transformations $\delta_L N=
(\epsilon_L^\alpha Q_{L \alpha}+\bar \epsilon_L^\alpha \bar Q_{L\alpha})N$,
$$
\delta_L X= (\epsilon+\bar \epsilon)_L^\alpha(Q+\bar Q)_{L\alpha} X-
(\epsilon-\bar \epsilon)_L^\alpha (Q-\bar Q)_{L \alpha} Y
$$
$$
\delta_L Y= (\epsilon+\bar \epsilon)_L^\alpha(Q+\bar Q)_{L\alpha} Y 
+(\epsilon-\bar \epsilon)_L^\alpha(Q-\bar Q)_{L\alpha} X
$$
and similarly for the right-moving supersymmetries. The combination 
$(\epsilon +\bar \epsilon)_L$ parameterizes a supersymmetry which is 
realized linearly on $X$ and $Y$ separately while 
$(\epsilon -\bar \epsilon)_L$ mixes the two. Therefore, $X$ and $Y$ are 
${}^N d= {}^2 3$ superfields.
We will henceforth use $X$ and $Y$ to denote the $\circ$-real and 
$\circ$-imaginary parts of a chiral superfield $N=M$. Similarly 
$S={\rm Re}_\circ H$ and $T={\rm Im}_\circ H$ will denote the 
half-supersymmetric projections of the twisted-chiral superfield  $H$.

It is easy to show by covariant projection that the ${}^N\!d={}^23$ 
superfields $X, S$ and $Y, T$ are strengths for a (partially on-shell) 
vector multiplet and a scalar multiplet respectively. 
For example for the field strength components of $X$ and $Y$ we find 
$$
\nabla_{L\alpha} \nabla_{R\beta}X|= (\gamma^{mn})_{\alpha \beta} F_{mn}
$$
$$
\nabla_{L\alpha}\nabla_{R\beta} Y|=(\gamma^m)_{\alpha \beta}\left(\partial_y A_m-\partial_m a\right)-i\varepsilon_{\alpha \beta}D
$$
$$
\nabla_{L\alpha} \bar \nabla_{R\beta}S|= (\gamma^{m})_{\alpha \beta} \tilde H_m-\varepsilon_{\alpha \beta}\partial_y l
$$
$$
\nabla_{L\alpha}\bar \nabla_{R\beta} T|=
(\gamma^m)_{\alpha \beta}\partial_m l+\varepsilon_{\alpha \beta}\tilde H_y
$$
where we have retained the $\partial_y$-terms for use in section 4.2 where 
we will interpret them as fluxes coupling to space-time 
defects.\foot{If a higher-dimensional theory is written in a 
lower-dimensional superspace notation then it is often the case that 
the `scalar' field strengths in the extra directions are related by 
equations of motion to $F$- and $D$-terms. For example, in six-dimensional 
SYM the $D$-term of the gauge field $V$ is related to the flux in 
the 5- and 6-directions by the equation of motion $D=F_{56}$ 
\refs{\MSS,\ArkaniHamedTB}
It is therefore possible to turn on these particular fluxes without 
breaking supersymmetry if we simultaneously give vevs of the 
same magnitude to these auxiliary terms. 
}

In the basis defined by 
$\theta^+_{\alpha}=(\theta_{L\alpha}+ \theta_{R\alpha})$ and 
$\theta^-_{\alpha}=(\theta_{L\alpha}- \theta_{R\alpha})$, $N=1$ 
decompositions may be represented (up to certain auxiliary fields) as 
$X=\Phi(\theta^+)+\theta^{- \alpha} W_\alpha(\theta^+)$ where $\Phi$ and 
$W_\alpha$ are the standard ${}^Nd ={}^13$ field strengths for a scalar 
and vector multiplet while $Y=\Phi^+(\theta^+)+\Phi^-(\theta^-)$ is the 
direct sum of two scalar superfields. 
The analogous statements hold for $S$ and $T$ in the basis 
$\theta^+_{\alpha}=(\theta_{L\alpha}+\bar \theta_{R\alpha})$, 
$\theta^-_{\alpha}=(\theta_{L\alpha}-\bar \theta_{L\alpha})$. In this 
case, the field strength $\Upsilon_\alpha$ analogous to $W_\alpha$ is 
the $d=3$ version of a variant representation of the tensor 
multiplet. 

Finally, let us comment very briefly on the structure of a real scalar prepotential $U$. 
As mentioned in section 2.3, the gravitational multiplet is 
represented such a field. As described in detail in 
reference \Eff, it has a gauge invariance 
$\delta U=\nabla^2_L \Lambda_L+\bar \nabla_L^2 \bar \Lambda_L+
\nabla_R^2 \Lambda_R+\bar \nabla_R^2 \bar \Lambda_R$ and can be put 
into the Wess-Zumino gauge
\eqn\Ufield{U=(h_{mn}+b_{mn}+l^{+-} \eta_{mn})\s^m_{\a\ad} \s^n_{\b\bd}\t_L^\a\tb_L^\ad
\t^\b_R{\bar \theta}_R^\bd+\dots}
where the ellipsis denotes higher-dimensional fields which will not enter our considerations.
Upon reduction under $\circ$, it is easy to check that ${\rm Re}_\circ U$ contains $h_{mn}$, $b_{mn}$, and $h_{yy}$ and ${\rm Im}_\circ U$ contains $h_{my}$, $b_{my}$, and $l^{+-}$. This result will be important when we discuss warping in section 4.

\subsec{$G_2$ Structure and ${\Bbb Z}_2$ Quotient}

The formalism developed in section 2 has $N=4$ supersymmetry in three dimensions. 
There are many ways to obtain a theory with a smaller amount of supersymmetry. 
Given an initial setup in which the compactification 
manifold is of the form ${\rm CY}_3\times S^1$, the obvious way to break 
half of the supersymmetry is by a $\Bbb Z_2$ quotient, which is a well-known 
way to obtain $G_2$ holonomy manifolds. The 
resulting spectrum is equivalent to a direct type IIA reduction 
on a general $G_2$ holonomy manifold. This example breaks 
supersymmetry within left- and right-moving sectors as we discuss presently.
A second way to reduce the amount of supersymmetry breaks 
between the two sectors by the introduction of some flux. 
We demonstrate the effects of such background fluxes in four dimensions in
section 4 by explicit worldsheet 
computations.

Our starting point is a Calabi-Yau 3-fold ${\rm CY_3}$ with complex 
structure $J$, symplectic $(1,1)$-form $\omega$, and holomorphic volume 
$(3,0)$-form $\Omega$. The Hodge structure is the familiar one with 
variable $h^{1,1}$ and $h^{2,1}$. 
Consider a conjugation acting freely on the Calabi-Yau defined such that 
$J\to -J$, $\omega\to -\omega$, and $\Omega\to \bar \Omega$. We extend this 
action to the circle with coordinate $y\in(-\pi, \pi]$ by the reflection 
$y\mapsto -y$. Let us denote the combined operation by 
$\sigma: {\rm CY}_3\times S^1 \to {\rm CY}_3\times S^1$. Although $\sigma$ 
fixes $y=0$ and $y=\pi$ on $S^1$, its action on ${\rm CY}_3$ is free and 
therefore the quotient $X=\left( {\rm CY}_3\times S^1\right)/ \sigma$ is 
smooth. The 3-form $\Phi=\omega \wedge {\rm d}y+{\rm Re}\Omega$, being 
invariant under the action of $\sigma$, descends to $X$ providing it with 
a $G_2$-structure \ref\Joyce{D.~D. Joyce. {\it Compact Manifolds with 
Special Holonomy}. Oxford Science Publications, Oxford, UK, (2000).}. 

Under the action of $\sigma$ the cohomology of the Calabi-Yau descends to 
the following. Since the space of 2-forms is real, it splits as 
$H^{1,1}=H^{1,1}_+\oplus H^{1,1}_-$ where the $H^{1,1}_\pm$ eigenspaces 
have the indicated eigenvalues. The odd forms are reflected through the 
vertical of the Hodge diamond since the involution acts as a conjugation. 
Therefore, the eigenspaces are subspaces of the sums $H^{3,0}\oplus H^{0,3}$,
spanned by ${\rm Re}\Omega$ and ${\rm Im}\Omega$,
and $H^{2,1}\oplus H^{1,2}$. Quotienting by $\sigma$ projects out the odd 
eigenspaces. The resulting cohomology of $X$ is real and has Betty 
numbers $b^2= h^{1,1}_+$ and $b^3=h^{1,1}_-+h^{2,1}+1$ where the $h^{1,1}_-$ 
terms in $b^3$ come from wedging with the 1-form ${\rm d}y$ on the circle.

We would like to extend this conjugation to the full superspace. Such an 
extension must be symmetric in its action on left-moving and right-moving 
fermionic coordinates.  From the discussion in section 3, the obvious 
candidate is the $\circ$-involution. Assuming this, we are in position to 
determine the spectrum. We take for definiteness type IIA on the Calabi-Yau 
3-fold. Then, as explained in section 2, the $h^{1,1}$ K\"ahler moduli are 
parameterized by the scalars in the ${}^N\!d={}^24$ vector multiplets $M_a$ 
while the $h^{2,1}$ complex moduli are embedded in the hypermultiplets 
$H_i$. Let us write the relevant vertex operators as
$$
\sum_{a=1}^{h^{1,1}} M_a  \Phi^a+ \sum_{i=1}^{h^{2,1}}H_i \Psi^i +{\rm h.c.}
$$
where the $\Phi^a$ generate $H^{1,1}$ and the $\Psi^i$ generate $H^{2,1}$. 
From the discussion above, we see that this 
expression decomposes under the extended conjugation as 
$$
\sum_{a=1}^{h^{1,1}_+} X_a \Phi_+^a
+\sum_{a=1}^{h^{1,1}_-} Y_a \Phi_-^a
+\sum_{i=1}^{h^{2,1}} S_i {\rm Re} \Psi^i
+{\rm h.c.}
$$
This result agrees with the standard component analysis
\refs{\PapadopoulosDA,\AcharyaGB}. With this construction on can take
a solvable model for the quotient 
$X=\left( {\rm CY}_3\times S^1\right)/ \sigma$ such as \roiban\ and 
use the hybrid formalism to compute supersymmetric amplitudes.

\newsec{Internal Ramond-Ramond Fluxes and Deformations}

We will now apply the methods of the hybrid formalism to 
the problem of internal Ramond-Ramond fluxes. Computations of superpotentials 
can be done by calculating scattering amplitudes with appropriate operators. 
In \LawrenceZK\ Lawrence and McGreevy have given a detailed discussion 
of the role of auxiliary fields and their meaning in terms of RR fluxes. We 
will use that analysis combined with the hybrid formalism to compute superpotentials and
effects due to torsion and warping.

It has been known since the beginning of the study of supersymmetric 
theories that giving vevs to auxiliary fields can be used to 
break supersymmetry. The connection between auxiliary fields and internal RR fluxes 
was made in \refs{\TaylorII,\VafaWI}. 

If $M_a(x,\th_L,\th_R)$ is a constant chiral superfield and there are
no fermionic background or Lorentz breaking terms, we have the simple form
$$M_a(\th_L,\th_R)=\phi_a+\th^2_L F_a+\th^2_R \bar F_a 
+\th_L^\alpha \th_R^\beta \e_{\a\b}D_a,$$
where $(F_a,\bar F_a,D_a)$ are auxiliary fields representing a combination 
of RR and NSNS fluxes \LawrenceZK. For general values of $(F,\bar F,D)$ 
supersymmetry is completely broken. There are two general cases 
preserving half of the supersymmetry. In the first one a combination of 
$(\t_L,\t_R)$ is preserved $\t_L+e^{i\gamma}\t_R$ where $\gamma$ is a phase 
depending on the specific model of supersymmetry breaking. In this case $M_a$ 
has the form 
\eqn\genedef{M_a=\phi_a + (\t_L+e^{i\gamma}\t_R)^2 F_a.}
In the second case the right- (or left-)moving supercharges are broken, preserving the other.
Then, $M_a$ takes the form
\eqn\semichiral{M_a=\phi_a+\t_R^2\bar F_a,}
where $F_a$ is due to NS flux.
In the case of $H_i$, the only combinations involving left- and right-moving 
product of $\t$ are vectors in space-time which means that $H_i$ can only be 
used to describe NSNS fluxes. 
$$H_i= \psi_i + \t_L^2 y_i,$$
where the left moving supersymmetry is broken. The physical meaning of 
all these auxiliary fields depends on the choice of type IIA or type 
IIB string as discussed in detail in \LawrenceZK. 

It was pointed out in \refs{\BerkovitsPQ,\LawrenceZK} that giving 
vevs to the auxiliary fields in compactification multiplets violates
the physical state conditions \physxi\ and \physsigma\ and that a possible 
solution to this apparent inconsistency is that the physical state conditions 
are modified when this type of flux is turned on. This means that 
giving vevs to those fields breaks $N=2$ superconformal invariance on 
the worldsheet. Instead of modifying the physical state {\it condition} 
we modify the physical {\it state} itself by 
observing that these deformations cannot be turned on alone. That is, there should 
be another deformation in the action that compensates the breaking due 
to non-zero values in the auxiliary components of $\Xi$ and $\Sigma$. 

Let us analyze one specific case to be more explicit. Suppose we want to add 
a deformation of the action corresponding to 
$\delta\Xi=(\t_L-\t_R)^2F_a\Phi_a$, breaking a combination of left-right 
moving supersymmetry. This vertex operator does not 
satisfy the physical state condition \physxi. A clear way to see this 
in terms of the familiar language of chiral states is that the vertex 
operator $(\t_L-\t_R)^2F_a e^{-\rho_L+\rho_R}\Omega^+_L(\Omega^+_R(\Phi^a))$ 
has a single pole with $G^+_L$ and $G^-_R$ viz. 
$$G^+_L[(\t_L-\t_R)^2F_a e^{-\rho_L+\rho_R}\Omega^+_L(\Omega^+_R(\Phi^a))]\to
{1\over z}e^{\rho_R}F_a\Omega^+_L(\Omega^+_R(\Phi^a)),$$ 
$$G^-_R[(\t_L-\t_R)^2F_a e^{-\rho_L+\rho_R}\Omega^+_L(\Omega^+_R(\Phi^a))]\to
{1\over {\bar z}}e^{-\rho_L}F_a\Omega^+_L(\Omega^+_R(\Phi^a)),$$
so it fails to be (anti-)chiral. (Note that 
$\CG^+_L[\Omega^+_L(\Omega^+_R(\Phi^a))]\to 0$
and $\CG^+_R[\Omega^+_L(\Omega^+_R(\Phi^a))]\to 0$.) To remedy this, one 
has to remember that one of the effects of fluxes is to generate 
torsions \grana\  and it turns out that at {\it linearized level in the 
deformation} semi-chiral and non-chiral (depending on the type of flux) 
operators have to be included in the internal CFT. These new operators are not 
physical by themselves either; only the combination is a consistent 
deformation of the background. This is the worldsheet counterpart to the 
target space result that torsions modify the closure conditions on the forms. 
A consequence of this in the hybrid formalism is that 
the theory does not factorize into two independent CFTs; space-time and 
internal superconformal generators are not conserved separately. Of course, the 
introduction of fluxes does not add new states in the spectrum implying
that these new semi-chiral and non-chiral vertex operators should 
not be independent. The physical operator will have the form 
\eqn\physicalrr{e^{-\rho_L+\rho_R}\Omega^+_L(\Omega^+_R(\delta \Xi))+
e^{\rho_R}W_1+e^{-\rho_L}W_2,}
where $e^{\rho_R}W_1+e^{-\rho_L}W_2$ is 
a vertex operator representing the effect of torsion with 
charge $(1,-1)$ and we are assuming it 
does not depend on space-time and $\t$. The (anti-)chirality conditions 
are now
\eqn\newchiral{0=G^+_L\left[(\t_L-\t_R)^2F_a 
e^{-\rho_L+\rho_R}\Omega^+_L(\Omega^+_R(\Phi^a))+
e^{\rho_R}W_1+e^{-\rho_L}W_2 \right]}
$$\to {1\over z}\left[F_ae^{\rho_R}\Omega^+_L(\Omega^+_R(\Phi^a)) + 
e^{\rho_R}\CG^+_L(W_1) +e^{-\rho_L}\CG^+_L(W_2)\right],$$
\eqn\newchiraltwo{0=G^-_R\left[(\t_L-\t_R)^2F_a 
e^{-\rho_L+\rho_R}\Omega^+_L(\Omega^+_R(\Phi^a))+
e^{\rho_R}W_1+e^{-\rho_L}W_2 \right]}
$$\to {1\over z}\left[F_a e^{-\rho_L}\Omega^+_L(\Omega^+_R(\Phi^a)) + 
e^{-\rho_L}\CG^+_R(W_2) +e^{+\rho_R}\CG^+_R(W_1)\right],$$
whence we obtain four equations determining $W_1$ and $W_2$:
$$F_a\Omega^+_L(\Omega^+_R(\Phi^a)) =-\CG^+_L(W_1),\quad\quad \CG^+_R(W_1)=0,$$
$$F_a\Omega^+_L(\Omega^+_R(\Phi^a)) =-\CG^+_R(W_2),\quad\quad 
\CG^+_L(W_2)=0.$$

In the large radius limit $\CG^+_L$ acts as $dy^I\partial_I$ and 
$\CG^+_R$ acts as $d\bar y^{\bar I}\partial_{\bar I}\,$ in our notation 
and the above equations are recognizable as the equations relating 
components of the intrinsic torsion to the 
un-deformed forms in the Calabi-Yau \grana. It should be stressed that we are 
considering only the first order in $F_a$. In the case of $\delta\Xi$, 
$(W_1,W_2)$ is a pair of semi-chiral and semi-anti-chiral vertex operators.
If \semichiral\ is used instead to deform 
the action, only $W_2$ would be needed and the deformed 
compactification manifold will not be complex.
Since $(W_1,W_2)$ are not chiral primaries, 
it not clear how to construct them in terms of operators corresponding to 
geometric objects. Nevertheless, at least in the classical limit, it 
should be possible to write a $\sigma$-model action including all possible 
corrections in $F_a$. This gives an exact form for $G^\pm_{L,R}$ and hence 
exact equations analogous to \newchiral\ and \newchiraltwo\ for $(W_1,W_2)$.

It is interesting to note that the combination 
$e^{-\rho_L+\rho_R}\Omega^+_L(\Omega^+_R(\delta \Xi))+e^{\rho_R}W_1+
e^{-\rho_L}W_2$ 
resembles the holomorphic ``three-form 
superfield'' proposed in \LawrenceZK\ and further discussed in \ref\GranaBG{
  M.~Grana, R.~Minasian, M.~Petrini and A.~Tomasiello, {\it Supersymmetric 
backgrounds from generalized Calabi-Yau manifolds,} JHEP {\bf 0408}, 046 (2004)
[arXiv:hep-th/0406137].}. It is likely that the non-linearized version 
of this vertex operator should be written using {\it pure spinors} \GualtieriDX, which arise naturally in the description of generalized compactifications.\foot{We would like to avoid a potential confusion with the pure spinor formalism of Berkovits \berk. However, although there is currently no explicit relation known between the Berkovits formalism 
and the {\it pure spinors} mentioned here, it is likely that they are intimately connected via the superconformal extension of the pure spinor string \pureconf\ compactified to four dimensions.}
 
Let us see how $(W_1,W_2)$ appear in the full CFT. If the 
compactification manifold is an exact Calabi-Yau, the vertex 
operators $\Phi^a$ and $\Psi^i$ can be written as 
$$\Phi^a=\omega_{\bar I J}^a\bar\psi_L^{\bar I}\psi_R^{J},\quad 
\Psi^i=g_{K\bar I}h^i{}^{K}_{\bar J}\bar\psi_L^{\bar I}\bar\psi_R^{\bar J},$$
where $(\psi_L,\psi_R)$ are the usual RNS fermions, 
$\omega^a_{I\bar J}$ is a harmonic $(1,1)$-form, $g_{I\bar J}$ is the 
Calabi-Yau metric and $h^i{}^{\bar K}_J$ is an element of the Dolbeault 
cohomology group $H^{1,0}(\bar T)$ of the Calabi-Yau. The introduction of fluxes deforms the 
original manifold and new cubic operators 
in the fermions
\eqn\newcftop{\Upsilon=u_{I,\bar J\bar K}
\psi_L^{I}\bar\psi_R^{\bar J}\psi_R^{\bar K},\quad 
\Theta= t_{I J,\bar K}\psi_L^{I} \psi_L^{J}\bar\psi_R^{\bar K},}
$$\bar \Upsilon=\bar u_{\bar I, JK}
\bar\psi_L^{\bar I}\psi_R^{J}\psi_R^{K},\quad 
\bar \Theta=\bar t_{\bar I\bar J, K}
\bar\psi_L^{\bar I}\bar\psi_L^{\bar J}\psi_R^{K},$$
 should be included.\foot{This choice is the most convenient for 
the discussion above. As in the case of the original chiral primaries, the 
cubic operators can be rotated using $(J^{++}_{L,R},J^{--}_{L,R})$.}

We must now determine the relation 
between $(u_{I,\bar J\bar K},\bar u_{\bar I, JK}, 
t_{IJ,\bar K},\bar t_{\bar I\bar J, K})$ and the original physical 
deformations in terms of the RR fluxes. 
From \newcftop\ we can select the candidates for $(W_1,W_2)$ 
by counting charges and assuming that 
$(\Upsilon,\bar\Upsilon,\Theta,\bar\Theta)$ are constant superfields in 
space-time;
\eqn\wansatz{W_1= \Upsilon=u_{I,\bar J\bar K} 
\psi_L^{I} \bar\psi_R^{\bar J}\bar\psi_R^{\bar K} ,\quad 
W_2=\Theta=t_{IJ,\bar K} \psi_L^{I}\psi_L^{J}\bar\psi_R^{\bar K},}
have the correct conformal weight. Substituting these into
\newchiral\ we have 
\eqn\torsioneq{ G^+_L ( 
e^{\rho_R}u_{I,\bar J\bar K}\psi_L^I\bar\psi_R^{\bar J}\bar\psi_R^{\bar K} +
e^{-\rho_L}t_{IJ,\bar K} \psi_L^{I}\psi_L^{J}\bar\psi_R^{\bar K} +} 
$$+ e^{-\rho_L+\rho_R}(\t_L-\t_R)^2 F_a\tilde\omega^a_{KM\bar K\bar M}
\psi_L^M\psi_L^K\bar\psi_R^{\bar K}\bar\psi_R^{\bar M} ) \to$$
$${e^{-\rho_L}\over z}(\partial_M t_{IJ,\bar K})
\psi_L^M\psi_L^I\psi_L^J\bar\psi_R^{\bar K}+ 
{e^{\rho_R}\over z}( \partial_M u_{K,\bar M\bar K} +
F_a\tilde\omega^a_{KM\bar K \bar M})
\psi_L^M\psi_L^K\bar\psi_R^{\bar M}\bar\psi_R^{\bar K}=0,$$
where $\tilde\omega^a_{KM\bar K\bar M}=
\Omega_{MK}{}^{\bar I}\bar\Omega_{\bar M\bar K}{}^J \omega^a_{\bar I J}$.
Similarly, from the condition \newchiraltwo\ we obtain
\eqn\torsioneqtwo{{e^{\rho_R}\over {\bar z}}(-\bar\partial_{\bar M} 
u_{I,\bar J\bar K})
\psi_L^I\bar\psi_R^{\bar M}\bar\psi_R^{\bar J}\bar\psi_R^{\bar K}+ 
{e^{-\rho_L}\over {\bar z}}( \bar\partial_{\bar M} t_{K M,\bar K} +
F_a\tilde\omega^a_{KM\bar K \bar M})
\psi_L^M\psi_L^K\bar\psi_R^{\bar M}\bar\psi_R^{\bar K}=0.}
Using  $\bar\partial=d\bar y^{\bar I}\partial_{\bar I}$ and
$\partial=dy^I\partial_I$ we can write these equations as
$$\partial t=\bar\partial u=0$$
$$\partial u =\bar\partial t = -F_a \tilde \omega^a$$
where $\tilde\omega^a\in H^{2,2}$ is the Hodge dual of 
$\omega^a\in H^{1,1}$ or, equvalently,
$$d(t-u)=0\quad,\quad
d(t+u)=-2F_a \tilde \omega^a.$$

In the absence of flux $\Upsilon$ and $\Theta$ 
should be identified with ($su(2)$ rotations of) one of the original vertex operators $\Psi^i$. 
Note that since the value of $F_a$ is quantized there is no modulus
corresponding to \physicalrr. In other words, it is a deformation of the 
$\sigma$-model preserving the full superconformal invariance but there is no 
massless space-time field corresponding to it. 
The next question to be addressed concerns how the presence of this 
vertex operator in the action affects the 
equations of motion for other modulus fields. One should expect that 
the equations of motion will show that with appropriate flux all the initial 
modulus fields will turn out to be massive. 

From the discussion above we see that the correct vertex operators 
describing the presence of fluxes in the compactification-dependent sector have 
the general form
\eqn\generalflux{
{\cal F}_{cc}=e^{-\rho_L}\Gamma + e^{-\rho_R}\Lambda + 
e^{-\rho_L-\rho_R}\Omega^+_L(\Omega^-_R(\Psi^i))(
\theta^2_L y_i +\bar\theta^2_R \bar y_i),
}
$$
{\cal F}_{ca}=e^{-\rho_L}\Theta+ e^{\rho_R}\Upsilon +
e^{-\rho_L+\rho}\Omega^+_L(\Omega^+_R(\Phi^a))(
\theta^2_L F_a + \theta^2_R \bar F_a + \theta_L\theta_R D_a),
$$
where $\Gamma=g_{I,JK}\psi_L^I\psi_R^J\psi_R^K$ and 
$\Lambda=l_{IJ,K}\psi_L^I\psi_L^J\psi_R^K$ are determined in terms of the 
fluxes $(y_i,\bar y_i)$ and the elements of $H^{1,0}(\bar T)$.
$({\cal F}_{ca},{\cal F}_{cc})$ are the operators corresponding to the superforms proposed in 
\refs{\LawrenceZK,\GranaBG} up to the chiral boson dependence which implements the
correct charges and conformal weights. 

Now let us see how the presence of flux affects the vertex operator 
corresponding to space-time deformations. Since linearized fluctuations 
in $U$ can only describe perturbations satisfying $R_{\mu\nu}=0$, the 
effects of fluxes cannot be seen as coupled equations of Ramond-Ramond 
operators and $U$ as in \generalflux. One could try to include 
\generalflux\ into the action and re-compute physical state conditions.
A more direct way is to compute UV divergences coming from interactions 
of ${\cal F}_{ca}$. For example, the composite operator 
${\cal F}_{ca}\bar {\cal F}_{ca}$ has a UV divergence
\eqn\divflux{ :{\cal F}_{ca}(z)\bar{\cal F}_{ca}(z): = {1\over {\epsilon^2}}
\t_L^\alpha{\t_R}_\alpha \bar\t_L^{\dot\alpha}\bar{\t}_{R\dot\alpha}
D_aD_{\bar b}{\goth g}^{a\bar b}+ \cdots}
where $\cdots$ contains terms with fewer $\t$s, 
${\goth g}^{a\bar b}$ is the Zamolodchikov 
metric\foot{This is one of the possible metrics. A second one can 
be defined as 
$\CG_L^+(\CG_R^+(\Phi^a))\CG_L^-(\CG_R^-(\bar\Phi^{\bar b}))\to 
{{\goth G}^{a\bar b}\over {|z-w|^4}}$, and is the metric of a different section 
of the moduli space \DixonBG.}  on the moduli space, and 
$\epsilon$ is a UV regulator in the OPE 
$$\Phi^a(z)\bar\Phi^{\bar b}(w)\to 
{{\goth g}^{a\bar b}\over { |z-w|^2+\epsilon^2}}~.$$


This divergence breaks conformal invariance. Because of its $\theta$ 
dependence, \divflux\ can only be cancelled by the vertex operator 
constructed from $U$ and since \generalflux\ does not break four-dimensional 
Poincar\'e invariance, this part of $U$ should be independent of $x$. As was mentioned 
in section 2.3, in a more general situation $U$ could be a function on
the compactification manifold. At one loop \divflux\ will be cancelled by 
$:T_L T_R U(\t_L,\t_R,\bar\t_L,\bar\t_R,y^I,\bar y^{\bar J}):$. This term 
is one of the many contributions from the integrated vertex operator \udef. In a flat ten-dimensional 
background the divergence coming from  this term is zero since it is 
proportional to $\Box_{10} U$ which vanishes for a massless deformation. In the 
case at hand, where $U$ is independent of $x$, we have 
\eqn\warping{ :T_L T_R U(\t_L,\t_R,\bar\t_L,\bar\t_R,y^I,\bar y^{\bar J}):
+:{\cal F}_{ca}(z)\bar{\cal F}_{ca}(z):=}
$$ {1\over \epsilon^2}\left( 
\Box_{\rm CY}U(\t_L,\t_R,\bar\t_L,\bar\t_R,y^I,\bar y^{\bar J}) + 
\t_L^\alpha{\t_R}_\alpha \bar\t_L^{\dot\alpha}\bar{\t}_{R\dot\alpha}
D_aD_{\bar b}{\goth g}^{a\bar b}\right)+\cdots =0.$$
The component $\t_L^\alpha{\t_R}_\alpha
\bar\t_L^{\dot\alpha}\bar{\t}_{R\dot\alpha}$ is precisely where the space-time 
metric sits in $U$ (see \Ufield) and equation \warping\ implies the usual space-time 
warping in flux compactifications. In \warping\ $\cdots$ include other 
divergent terms that should also be cancelled, and this implies further 
corrections to the original background. 

To compute superpotentials generated by fluxes in  
superstring scattering amplitudes we include \generalflux\ in the physical 
vertex operators \phymoduli\ with appropriate $su(2)$ rotations. For example, 
in the case of \genedef, there is a superpotential of the type
$${\cal W}=3\int d^4x d^2(\t_L-\t_R)F_a M_b M_c \CK^{abc},$$
where $d^2(\t_L-\t_R)$ is the measure for the preserved 
supersymmetry. Because of the mixing of $(\rho_L,\rho_R)$ and 
$(\CJ_L,\CJ_R)$ charges in \generalflux\ and the zero-mode measure \normfinal, 
the potential contributions 
from $\Theta$ and $\Upsilon$ do not appear and
the presence of flux does not modify the topological amplitudes as argued by Vafa in \VafaWI. It should be noted, however, that now $F_a$ is not interpreted 
as the auxiliary field in $M_a$, but a component of the physical 
operator ${\cal F}_{ca}$.
To further study the supersymmetry breaking in this case, one can use the superspace projection technique introduced in Section 3.

\newsec{Conclusions and Further Directions}

In this paper we analyzed the hybrid formalism for supersymmetry-breaking 
backgrounds. It was shown how worldsheet 
techniques can be used to study backgrounds with internal 
RR fluxes and a covariant formulation of the Type IIA string on 
$\left( {\rm CY}_3\times S^1\right)/ \Bbb Z_2$ was constructed. 
The significant simplification of calculations in the hybrid
formalism relative to the analogous ones in RNS 
is due to the absence of spin fields in the former. It was shown how 
the hybrid formalism can be applied to a much wider class of compactifications than
the original ${\rm CY}_3$ case; much remains to be done in this direction.

There are many interesting applications that can follow from this work. 
For example, the hybrid is suitable for describing intersecting brane models 
which are currently one of the possible approaches to string phenomenology. 
Here the substitution of supercharges in favor of spin fields simplifies, 
among other things, the 
calculation of correlation functions associated to 
higher-dimension operators in the effective theory. 

In a closely related line of research, one can use the 
hybrid formalism to study flux compactifications on Calabi-Yau orientifolds 
using a generalization of the superspace reduction technique of section 3. 
For example, in the Type IIA case, the inclusion of O6 planes projects 
out half of the spectrum by a superspace involution switching 
$\theta_L\leftrightarrow \theta_R$ in the case of chiral fields and 
$\theta_L\leftrightarrow \bar \theta_R$ in the case of twisted-chiral 
fields. Similarly to the analysis of section 4.1, the surviving 
half-supersymmetric ${}^N\!d={}^14$ superfields are correlated with the 
induced projection on cohomology.

Many tools have been developed to study strings on Calabi-Yau spaces, like 
Gepner models \ref\GepnerVZ{
  D.~Gepner,{\it Exactly Solvable String Compactifications On 
Manifolds Of SU(N) Holonomy,} Phys.\ Lett.\ B {\bf 199}, 380 (1987).}, linear 
$\sigma$-models \ref\WittenYC{E.~Witten, {\it Phases of N = 2 theories in 
two dimensions,} Nucl.\ Phys.\ B {\bf 403}, 159 (1993) 
[arXiv:hep-th/9301042].}, and topological strings \ref\NeitzkeNI{A.~Neitzke 
and C.~Vafa,{\it Topological strings and their physical applications,}
arXiv:hep-th/0410178.}. The presence of fluxes modifies the compactification 
CFT and such tools are no longer suitable. Since the main ingredient of these 
techniques is the $N=2$ supersymmetry 
algebra on the worldsheet and as we have shown that this symmetry is preserved, 
it is possible that there exist generalizations 
of these methods. For example, the $\hat c=5$ formalism \BerkovitsPQ\ has a 
linearly realized supersymmetry algebra on the worldsheet and one can add the 
space-time sector and a linear $\sigma$-model describing the internal space. One then searches
for actions where the two algebras do not  decouple. This would correspond 
to a generalized compactification. Another possibility is that topological 
strings on generalized complex spaces recently considered by Pestun in 
\ref\PestunRJ{V.~Pestun,
{\it Topological strings in generalized complex space,} arXiv:hep-th/0603145.}
will play an important role in future studies. 
\lref\KappeliFJ{
  J.~Kappeli, S.~Theisen and P.~Vanhove,
  {\it A note on topological amplitudes in hybrid string theory,}
  arXiv:hep-th/0607021.
}
Quantum corrections, both string loop and $\alpha'$, should also be 
considered. Loop amplitudes of compactification-dependent states could be calculated in a 
supersymmetric way as in \refs{\onehyb,\KappeliFJ}. In order to pursue this, one has to understand better the
correlation functions of the time-like chiral boson $\rho$. To go beyond 
the linearized level, a general hybrid $\sigma$-model action with fluxes and 
warping can be constructed along the lines of \Eff. This 
$\sigma$-model action is equivalent to the action of the uncompactified 
ten-dimensional superstring using $d=4$ $N=2$ notation. This would allow us 
to compute $\alpha'$ corrections and consistency conditions for 
backgrounds with warping and internal RR flux using, for example, the beta 
function method. We hope to give simple examples of the hybrid formlism in
this type of backgrounds such as D$p$-brane solutions in the future.

\vskip 15pt
{\bf Acknowledgments:} We would like to thank Nathan Berkovits and
Ram Sriharsha for discussions and comments. The work of BCV 
is supported by NSF grant number PHY 0354776 
and the work of WDL3 is supported by NSF grant numbers PHY 0354776 and DMS 0502267. 
BCV also thanks University of North Carolina at Chapel Hill, where a part 
of this work was done, for their hospitality.

\appendix{A}{Hybrid Compactified on $S^1$: Supersymmetric Operators in $d=3$}

In this appendix we write, in a convenient way, the hybrid variables for the compactification Type II strings on ${\rm CY}_3\times S^1$.
With the definitions of section 3, the supersymmetric 
operators are
\eqn\newmomen{\Pi^{\a\b}_L=\p_L x^{\a\b}+ \t_L^{(\a}\p_L \tb^{\b)}_L +
\tb_L^{(\a}\p_L \t^{\b)}_L, \quad Z=\p_L x_2 +\t_L^{\a}\p_L \tb_{L\a } +
\tb_L^{\a}\p_L \t_{L\a}} 
$$ d_{L\a}= p_{L\a}+\tb_L^\b \p_L x_{\a\b} -\half \tb^2\p_L\t_{\a} + 
{1\over 4} \t_{L\a}\p_L \tb^2 +\p_L x_2 \tb_\a,$$
$$ \db_{L\a}=\bar p_{L\a}+ \t_L^\b\p_L x_{\a\b} -\half \t_L^2\p_L\tb_{L\a} + 
{1\over 4} \tb_{L\a}\p_L \tb_L^2 +\p_L x_2 \t_{L\a},$$
where $Z$ is a supersymmetric extension of the central charge operator $\partial_L x_2$.
The algebra of these operators is
\eqn\newalge{d_{L\a} \bar d_{L\b}\to {1\over{z-w}}(\Pi_{L\a\b}+ 
\e_{\a\b}Z),\quad d_{L\a}d_{L\b}\to 0, \quad \bar d_{L\a} \bar d_{L\b}\to 0,}
$$d_{L\a}\Pi_{L\b\g}\to {1\over{z-w}}\e_{\a(\b}\p_L\tb_{\g)},\quad 
\bar d_{L\a}\Pi_{L\b\g}\to {1\over{z-w}}\e_{\a(\b}\p_L\t_{L\g)},$$
$$d_{L\a}Z\to {1\over{z-w}}\p_L\tb_{L\a},\quad 
\bar d_{L\a}Z\to {1\over{z-w}}\p_L\t_{L\a}$$
With these definitions, the superconformal algebra is 
\eqn\threesup{T=\half \Pi_{L\a\b}\Pi_L^{\a\b}+\half Z_L^2+ d_{L\a}\p_L\t_L^\a +
\bar d_{L\a}\p_L\tb_L^\a +\CT_{CY},}
$$G^+=e^{\rho_L}d^2+\CG^+_{CY},\quad G^-=e^{-\rho_L}\bar d^2+ \CG^-_{CY}, 
\quad J=-\p_L\rho_L + \CJ_{CY}.$$
The spectrum is now characterized by the eigenvalues of the central charge 
operators if we compactify $x_2$ on a circle. Since there is left- and 
right-moving central charge, superfields are classified by two 
integers $(n,m)$ where $n$ is the Kaluza-Klein momentum 
and $m$ is the winding number. 

Since we are breaking the full SO(1,3) covariance, we can define new 
super-covariant operators like $:\!d\bar d\!:$, $d\p_L \tb$, 
and $\bar d\p_L\t$. The algebra that is generated by these operators is a 
higher-spin algebra. It is likely that after breaking space-time and 
worldsheet supersymmetry these operators are related to the $G_2$ holonomy 
conformal algebra found by Shatashvili and Vafa \shava\ for a more general 
$G_2$ manifold than the special case considered here.

\listrefs

\end